\documentclass[a4paper,12pt]{article}
\usepackage{amsmath}
\usepackage{amssymb}
\usepackage[latin1]{inputenc}
\usepackage{amsfonts}
\usepackage{graphics}
\usepackage{epsfig}
\usepackage{epic}

\begin{document}
\begin{center}
{\large THE MASSES OF GAUGE FIELDS IN HIGHER\\ SPIN FIELD THEORY ON ADS(4)}\\
\vspace{2cm}
{\large W. Rühl}\\
Department of Physics, Kaiserslautern University of Technology,\\P.O.Box 3049,
67653 Kaiserslautern, Germany \\ 
\vspace{5cm}
\begin{abstract}
Higher spin field theory on AdS(4) is defined by lifting the minimal conformal 
sigma model in three dimensional flat space. This allows to calculate           the masses from the anomalous dimensions of the currents in the sigma model.
The Goldstone boson field can be identified. 
\end{abstract}
\vspace{5cm}

{\it August 2004}
\end{center}
\newpage

\section{Introduction}
We consider the higher spin field theory constructed on AdS(4)space by lifting
of the O(N) vector minimal sigma model from R(3)[1]. It results a nonlocal 
interacting renormalized QFT denoted HS(4) and defined by its n-point functions which can each be computed by a 1/N expansion. Such AdS field theory has the 
unconventional property that mass renormalizations of the fundamental fields 
and composite fields can be calculated perturbatively. This is made possible 
by the fact that the corresponding flat conformal field theory permits to 
calculate the anomalous part $\eta$ of the conformal dimension $\Delta$ by 1/N 
expansion and that the mass $m^{2}$ of the AdS field and $\Delta$ of the dual 
conformal field are related, e.g. for symmetric tensor fields of rank $l$ by 
\begin{equation}
m_{l}^{2} = \Delta(\Delta-d) - (l-2)(d+l-2)
\label{1}
\end{equation}
Obviously for conserved currents $J^{(l)}(x)$ with exact dimension
\begin{equation}
\Delta(l) = d+l-2
\label{2}
\end{equation}
we obtain a vanishing mass corresponding to the gauge fields $h^{(l)}$
on AdS space.

Lifting of the flat conformal field theory is done with the bulk-to-boundary 
propagator (say, in the scalar case, see [2] for the general case) 
\begin{equation}
K_{\Delta}(z, \vec x)= \left(\frac{z_0}{z_{0}^{2}+ (\vec z - \vec x)^{2}}
\right)^\Delta\\
(z \in AdS(4), \vec x \in R(3)) 
\label{3}
\end{equation}
which satisfies the free field equation
\begin{equation}
(D_z -m^{2})K_\Delta(z,\vec x) = 0 
\label{4}
\end{equation}
with $m^{2}$ as in (1) and $D_z$ a covariant second order differential 
operator. It is this equation which transforms dimension into mass. 

If the symmetric tracesless tensor current $J^{(l)}(\vec x)$ assumes an 
anomalous dimension $\eta(l)$ (which it does for $l\geq4$)
\begin{equation}
\Delta(l) = d+l-2+\eta(l)
\label{5}
\end{equation}
which can be 1/N expanded as 
\begin{equation}
\eta(l) = \sum_{r=1}^{\infty} \frac{\eta_{r}(l)}{N^r}
\label{6}
\end{equation}
we obtain masses for $h^{(l)}$ from (1) and (5)
\begin{equation}
m_{l}^{2} = \eta(l)[d+2(l-2)] + \eta(l)^{2} 
\label{7}
\end{equation}

M. Porrati [3] has shown that a Higgs phenomenon is possible for a graviton
to assume a mass. A massless symmetric tensor representation of the 
conformal group $[\Delta,l]$ with
\begin{equation}
\Delta = d-2+l|_{d=3} = l+1
\label{8}
\end{equation}
which is massless, appears from an irreducible massive representation
in the limit 
\begin{equation}
\lim_{\Delta\to l+1} [\Delta,l] = [l+1,l]\oplus[l+2,l-1]
\label{9}
\end{equation} 
The second representation $[l+2,l-1]$ is identified with the Goldstone field. 
L. Girardello et al.[4] propose to create the Goldstone field from tensoring 
the conserved current $[l-1,l-2]$ corresponding to $J^{(l-2)}$ or $h^{(l-2)}$
with the scalar field $\alpha(\vec{x})$ on R(3) resp. $\sigma(z)$ on AdS(4)
of dimension two (all in the free field limit)
\begin {equation}
[l-1,l-2]\otimes [2,0] = \bigoplus_{s=0}^{\infty}\bigoplus_{n=0}^{\infty}
[l+s+n+1, l+s-2]
\label{10}
\end{equation}
which contains the Goldstone field representation for $s = 1$, $n = 0$. 
In the minimal sigma model the operator product of the current $J^{(l)}$
with the scalar field $\alpha$ contains by expansion the currents 
$J^{(l,t)}$ with dimension $d-2+l+2t$, $t \in\mathbf{N}$, (in the free field
limit). Thus the Goldstone field of $J^{(l)}$ is $J^{(l-1,1)}$. A Higgs
mechanism for producing the masses of the gauge fields is possible therefore
except for the graviton since in (10) $l=2$ is excluded (the representation 
$[1,0]$ does not occur but is eliminated in favour of the dual representation
$[2,0]$ by the boundary condition of AdS(4)).

The gauge fields contain a traceless symmetric part $h^{(l)}$ and companion 
fields that are dynamically irrelevant. In HS(4) there exists a bilocal
biscalar field $B(z_1,z_2)$ [1] which by operator product expansion decomposes 
into all $h^{(l)}$, $l \in 2\mathbf{N}$, and the scalar field $\sigma(z)$ at
leading order in N, and further operators at order 1/N and higher.
The starting point for a calculation of the masses $m_{l}^{2}$ is therefore the 
AdS four-point function
\begin{equation}
<B(z_1,z_3), B(z_2,z_4)>_{AdS}  
\label{11}
\end{equation}

\section{The perturbative corrections to the bilocal fields}
Instead of the AdS Green function (11) we study the flat CFT Green 
function
\begin{equation} 
<b(\vec x_{1},\vec x_{3})b(\vec x_{2},\vec x_{4})>_{CFT}
\label {12}
\end{equation}
where $b$ is defined from the Lorentz scalar O(N) vector fields $\vec\varphi(
\vec x)$ by the normal product [1]
\begin{equation}
b(\vec x_1,\vec x_3) = N^{-1/2}\vec\varphi(\vec x_1)\vec\varphi(\vec x_3)
\label{13}
\end{equation}
This O(N) vector field is normalized to
\begin{equation}
<\varphi_{i}(\vec x_1)\varphi_{j}(\vec x_2)> = \delta_{ij}(\vec x_{ij}^{2})^{-\delta} \qquad
i,j\in {1,2\ldots N}
\label{14}
\end{equation}
where $\delta$ is the conformal dimension of $\varphi$ (we set $d = 3$ at
 the end)
\begin{eqnarray}
\delta &=& \mu - 1 +\eta(\varphi), \qquad \mu = \frac{d}{2}
\label{15} \\
\eta(\varphi) &=& \sum_{r=1}^{\infty}\frac{\eta_{r}(\varphi)}{N^{r}}
\label{16}
\end{eqnarray}
The first three terms in (16) are known [5].

Due to the contraction of O(N) vector indices the $O(1)$ contribution
to the Green function is from the graphs $A_1$, $A_3$   
\vspace{0.5cm}
\begin{figure}[ht]
\centering
\includegraphics[angle=0,totalheight=3cm]{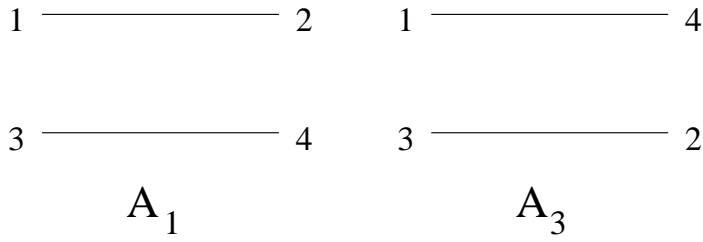}
\caption{\small The disconnected graphs $A_1$ and $A_3$}
\end{figure}
\vspace{0.5cm}
with the result
\begin{equation}
(\vec x_{12}^{2}\vec x_{34}^{2})^{-\delta} + (\vec x_{14}^{2}\vec x_{23}^{2})^
{-\delta}
\label{17}
\end{equation}
and from the exchange graph $B_2$ of the scalar field $\alpha$
with dimension $\beta$
\begin{equation}
\beta = 2 -2\eta(\varphi) -2\kappa
\label{18}
\end{equation}
Here $\kappa$, the ``conformal dimension of the coupling constant'' is of 
order $O(1/N)$ and can be expanded in powers  of 1/N (see A. N. Vasilev 
et al. [5]).
\vspace{0.5cm}
\begin{figure}[ht]
\centering
\includegraphics[angle=0,totalheight=3cm]{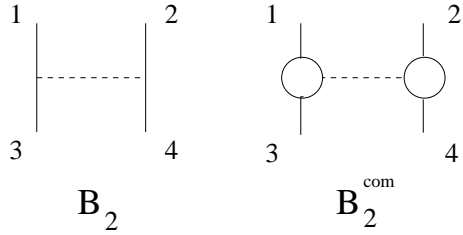}
\caption{\small The graph $B_{2}$ and its complete radiatively 
corrected form $B_{2}^{com}$}
\end{figure}
\vspace{0.5cm}
At leading order $B_2$ yields
\begin{equation}
(\vec x_{12}^{2}\vec x_{34}^{2})^{-\delta}\{C_1u^{\mu-2\delta}F_{\alpha}(u,v) +
C_2F_{\vec{\varphi}^{2}}(u,v)\}
\label{19}
\end{equation}
where the first (second) term describes the exchange of 
$\alpha$ $(\vec{\varphi}^{2})$, respectively, which are dual to 
each other in a representation
theoretic sense. Expressions in (19) are
\begin{eqnarray}
F_{\alpha}(u,v) &=& \sum_{n,m=0}^{\infty}\frac{u^{n}(1-v)^{m}}{n!m!}
\frac{(n!)^{2}((n+m)!)^{2}}{(2n+m+1)!(3-\mu)_n}
\label{20} \\
F_{\vec\varphi^{2}}(u,v)& =& \sum_{n,m=0}^{\infty}\frac{u^{n}(1-v)^{m}}
{n!m!}\frac{((\mu-1)_{n}(\mu-1)_{n+m})^{2}}{(2\mu-2)_{2n+m}(\mu-1)_{n}}
\label{21} \\
u &=& \frac{\vec x_{13}^{2}\vec x_{24}^{2}}{\vec x_{12}^{2}\vec x_{34}^{2}}
\label{22} \\
v &=& \frac{\vec x_{14}^{2}\vec x_{23}^{2}}{\vec x_{12}^{2}\vec x_{34}^{2}}
\label{23}
\end {eqnarray}
The constants $C_{1,2}$ involve the coupling constant $z$ squared between 
$\alpha$ and $\vec\varphi$ to leading order $z_1$
\begin{eqnarray}
z &=& \sum_{r=1}^{\infty}\frac{z_{r}}{N^{r}}
\label{24} \\
z_1 &=& 2\pi^{-2\mu}\frac{(\mu-2)\Gamma(2\mu-2)}{\Gamma(\mu)\Gamma(1-\mu)}
\label{25}
\end{eqnarray}
and are explicitly given as
\begin{eqnarray}
C_1 &=& \frac{\Gamma(2\mu-1)}{ \Gamma(\mu)^{2}\Gamma (1-\mu)\Gamma(\mu-1)(\mu-2)}
\label{26}\\
C_2 &=& -2
\label{27}
\end{eqnarray}

In CFT in flat space conformal invariance determines three-point functions
up to a few normalizing constants. The same must be true then for complete 
exchange graphs. For the radiatively corrected (at the vertices) graph 
$B_{2}^{com}$ of $B_{2}$ we obtain instead of (19),(21)
\begin{equation} 
C_{2}^{com}(\vec x_{12}^{2} \vec x_{34}^{2})^{-\delta}F_{\vec\varphi^{2}}^
{com}(u,v)
\label{28} 
\end{equation}
with
\begin{eqnarray}
C_{2}^{com} &=& -2 +O(1/N)
\label{29} \\
F_{\vec\varphi^{2}}^{com} &=& u^{\kappa}\sum_{n,m=0}^{\infty}\frac{u^{n}
(1-v)^{m}}{n!m!}\frac{((\Delta)_{n}(\Delta)_{n+m})^{2}}{(2\Delta)_{2n+m}
(2\Delta-\mu+1)_{n}}
\label{30} \\
\Delta &=&  \delta + \kappa
\label{31} 
\end{eqnarray}
where $\delta$ and $\kappa$ are taken from (15) and (18). The part O(1/N)
in $C_{2}^{com}$ can stay undetermined (see below).

In the subsequent section we shall decompose the O(1) contribution to the
Green function (12) into ``conformal partial waves'', amplitudes for the
exchange of irreducible conformal fields. The O(1/N) contribution to this
Green function contains three different types of terms:
1. Power series in $u$ and $1-v$ containing the same conformal partial 
waves as the order O(1). These imply a renormalization of the coupling
constants in the exchange amplitudes;
2. Power series in $u$ and $1-v$ multiplied with $logu$. Consistency 
requires that conformal partial wave expansion of the power series
yields the same exchange fields as the O(1) expansion. From the normalization 
we extract the combination
\begin{equation}
\frac{1}{2}(\eta(C) - \eta(A) -\eta(B))
\label{32}
\end{equation}
where the field C is exchanged and the fields A, B, C form a vertex of the 
exchange graph.
3. Power series in $u$ and $1-v$ containing new conformal partial waves. 

In the case of the Green function (12) the order O(1)
gives the exchanged fields $J^{(l)}, l\geq2$ even, and $\alpha$. At O(1/N)
there appear as new fields $J^{(l,1)}$, which contain the Goldstone fields. 
Since we are interested here only in the anomalous dimensions of $J^{(l)}$
it suffices to extract the $logu$ power series. These appear from the 
difference $B_{2}^{com} - B_{2}$ by expansion of the factor $u^{\kappa}$
and from four new graphs $B_1, B_3, C_{21}, C_{22}$, where the first two 
$B_{1,3}$ are obtained by crossing $B_2$, $C_{21}$ is a box graph, and the 
last one, $C_{22}$, is obtained by crossing $C_{21}$
\vspace{0.5cm}
\begin{figure}[ht]
\centering
\includegraphics[angle=0,totalheight=3cm]{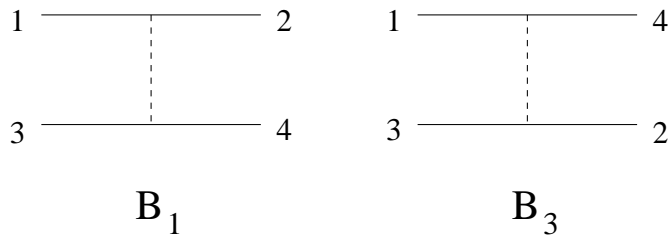}
\caption{\small The crossed exchange graphs.}
\end{figure}
\vspace{0.5cm}

The exchange graphs $B_{1} + B_{3}$ are easily calculated and give 
([6], equ. (3.17), only the $logu$ terms)
\begin{eqnarray}
& &N^{-1}\frac{\mu}{\mu -2}\eta_{1}(\varphi)(\vec x_{12}^{2}\vec x_{34}^{2})^
{-\delta}[-logu]\sum_{n,m=0}^{\infty}\frac{u^{n}(1-v)^{m}}{n!m!} \nonumber
\\ & & \left\{\frac{(\mu-1)_{n})(\mu-1)_{n+m}(n+m)!}{(\mu)_{2n+m}} 
+ \frac{((\mu-1)_{n+m})^{2}n!}{(\mu)_{2n+m}} \right\}
\label{33}
\end{eqnarray}
\vspace{0.5cm}
\begin{figure}[ht]
\centering
\includegraphics[angle=0,totalheight=3cm]{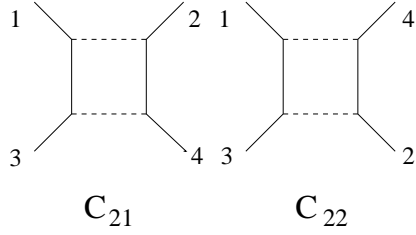}
\caption{\small The box graphs.}
\end{figure}
\vspace{0.5cm}
From the box graph  $C_{21}$ we get ([7], only the $logu$ terms)
\begin{eqnarray}
& &N^{-1}\frac{\mu}{\mu-2}\eta_{1}(\varphi)2(2\mu -3)( \vec x_{12}^{2}
\vec x_{34}^{2})^{-\delta}[-logu]\sum_{n,m=0}^{\infty}\frac{u^{n}(1-v)^{m}}
{m!}\nonumber \\
& &\frac{[(\mu-1)_{n+m}]^{2}}{(\mu)_{2n+m}}\sum_{p=0}^{n}\frac{(\mu-2)_{p}
(\mu-1)_{n+m+p}}{p!(2\mu-3)_{n+m+p}}
\label{34}
\end{eqnarray}
To obtain the box graph $C_{22}$ one can apply crossing explicitly by 
exchanging $\vec x_{2}$ with $\vec x_{4}$ or
\begin{eqnarray}
& &u \to\frac{u}{v} \qquad v \to\frac{1}{v}
\label{35} \\
& &(\vec x_{12}^{2}\vec x_{34}^{2})^{-\delta}u^{n}(1-v)^{m}[-logu]\rightarrow 
\nonumber \\
& &(\vec x_{12}^{2}\vec x_{34}^{2})^{-\delta}u^{n}(1-v)^{m}[-logu+logv]
\{(-1)^{m}\sum_{s=0}^{\infty}\frac{(n+m+\delta)_{s}}{s!}(1-v)^{s}\}
\label{36}
\end{eqnarray}
Inserting this into (34) we obtain the crossed graph contribution.

\section{Deriving the anomalous dimensions}
The Green function (12) can be submitted to a conformal partial wave 
decomposition by expressing it as a sum over exchange graphs (only the 
direct term) of $\alpha$ and $J^{(l)}$ at O(1), composites of two $\alpha$
and $J^{(l,1)}$ at O(1/N) and so on [8]. For  example we consider exchange 
graphs of $J^{(l,t)}$ with conformal dimension $\Delta(l,t)$
\begin{equation}
(\vec x_{12}^2\vec x_{34}^{2})^{-\delta}u^{t+\frac{1}{2}(\eta(l,t)
-2 \eta(\varphi))}\sum_{n,m=0}^{\infty}\frac{u^{n}(1-v)^{m}}{n!m!}\alpha_
{n,m}^{(l,t)}
\label{37}
\end{equation}
Here we use an adhoc normalization
\begin{equation}
\alpha_{0,l}^{(l,t)}= 1
\label{38}
\end{equation}
If the external legs of the four-point function are all equal, the
coefficients $ \alpha_{n,m}^{(l,t)}$ are independent of the external fields
(their dimension). Assume their dimension is $\delta$. For the 
currents $J{(l)}$ these coefficients are particularly simple (since they belong
to exceptional representations)
\begin{equation}
\alpha_{n,m}^{(l)} = \sum_{s=0}^{n}(-1)^{s}{n\choose s}{m+n+s \choose l}
\frac{(\delta+l)_{m-l+n}(\delta+l)_{m-l+n+s}}{(2\delta+2l)_{m-l+n+s}}
\label{39}
\end{equation}
These matrices  for $ n=0$ are easily inverted
\begin{equation}
\sum_{m=0}^{(l)}\beta_{l,m}\alpha_{0,m}^{(l')} = \delta_{l,l'}
\label{40}
\end{equation}
with
\begin{equation}
\beta_{l,m} = (-1)^{l-m}{l \choose m}\frac {((\delta+m)_{l-m})^{2}}
{(2\delta+m+l-1)_{l-m}}
\label{41}
\end{equation}
This inversion formula was given in [8],(4.11), (4.12) in a more general 
version with two free parameters instead of only one ($\delta$).

At order $O(1)$ we have the graphs $A_1+A_3+B_2$. We skip the $F_\alpha$
term and get for the remainder
\begin{equation}          
(\vec x_{12}^{2}\vec x_{34}^{2})^{-\delta}[1 +v^{-\delta} 
+C_{2}F_{\vec{\varphi}^{2}}] = \\
(\vec x_{12}^{2}\vec x_{34}^{2})^{-\delta}\sum_{m,n=0}^{\infty}
\frac {u^{n}(1-v)^{m}}{n!m!}a_{n,m}
\label{42}
\end{equation}
where
\begin{equation}
a_{n,m} = \delta_{n,0}(\delta_{m,0}+(\delta)_{m})\\
-2\frac{(\delta)_{n}((\delta)_{n+m})^{2}}{(2\delta)_{2n+m}}
\label{43}
\end{equation}
We recognize immediately that 
\begin{equation}
a_{0,0} = a_{0,1} = 0
\label{44}
\end{equation}
which implies that a current $J^{(l=0)}$ which would be identical with
$\vec\varphi^{2}$ is not exchanged. The ansatz
\begin{equation}
a_{n,m} = \sum_{l=2,even}^{\infty}\gamma_{l}^{2}\alpha_{n,m}^{(l)}
\label{45}
\end{equation}
is easily solved first by setting $n=0$ using the inversion formula 
(40) and then showing that it remains true for all $n$. The result is 
\begin{eqnarray}
\gamma_{l}^{2} &=& \frac {2((\delta)_{l})^{2}}{(2\delta-1+l)_{l}}\qquad
\mbox{for $l\geq$ 2, even} \nonumber\\
&=& 0 \qquad \mbox{for all other $l$}
\label{46}
\end{eqnarray}

At order $O(1/N)$ we obtain the relevant $logu$ terms from
\begin{equation}
B_1 + B_3 + C_{21} + C_{22}+ \{C_{2}^{com}F_{\vec\varphi^{2}}^{com}
-C_{2}F_{\vec\varphi^{2}}\}
\label{47}
\end{equation}
They sum up to 
\begin{equation}
N^{-1}(\vec x_{12}^{2}\vec x_{34}^{2})^{-\delta}\sum_{n,m=0}^{\infty}
\frac  {u^{n}(1-v)^{m}}{n!m!}[-b_{n,m}logu +c_{n,m}] 
\label{48}
\end{equation}
where $c_{n,m}$ is not known explicitly due to some nonevaluated integrals.
But the $b_{0,m}$ (it suffices to give these only for $n=0$) are  
\begin{eqnarray}
b_{0,m} &=& \frac{\mu(\mu-1)}{(\mu-2)[\mu-1+m]}\eta_{1}(\varphi)\nonumber\\
& &\{m! + (\mu-1)_{m} +2(2\mu-3)[\frac {((\mu-1)_{m})^{2}}{(2\mu-3)_{m}}
\nonumber\\
& &+m!\sum_{p=0}^{m}\frac{(\mu-1)_{p}(\mu-2)_{p}}{p!(2\mu-3)_{p}}]
-2(4\mu-5)\frac{(\mu)_{m}(\mu-1)_{m}}{(2\mu-2)_{m}}\}
\label{49}
\end{eqnarray}
Here we inserted $\kappa_{1}$, the coefficient of 1/N in the 1/N-expansion 
of $\kappa$, in the last term . But requiring consistence we must have 
\begin{equation}
b_{0,0} = b_{0,1}=0
\label{50}
\end{equation}
which implies the known result
\begin{equation}
\kappa_{1} = \frac{\mu}{\mu-2}\eta_{1}(\varphi)(4\mu-5)
\label{51}
\end{equation}
We recognize that the limit $d\to 3$ is rather delicate in (49) and that 
the point $d=3$ can only be reached by analytic continuation in $d$. 
 
We solve next 
\begin{equation}
b_{n,m} = \sum_{l\geq2, even}[\eta_{1}(\varphi)-½\eta_{1}(l)]\gamma_{l}^
{2}\alpha_{n,m}^{(l)}
\label{52}
\end{equation}
only for $n=0$, and find after some algebra with the inverting matrix 
$\beta$ (40), (41)
\begin{eqnarray}
\eta_{1}(2) &=& 0 
\label{53} \\
\eta_{1}(l) & = & \eta_{1}(\varphi)\frac{2(\mu-1)(2\mu+l-1)}{(2\mu-1)
(\mu+l-2)_{2}}
\{2(l-1) \nonumber 
\\ &+& \sum_{p=1}^{½l-2}((p+1)!)^{2}{l \choose p+1}\frac{(2\mu+1+p)_
{l-4-2p}}{(2\mu+1)_{l-4}}\}\qquad l\geq 4
\label{54}
\end{eqnarray}
Consistency requires that (52) is also solved for $n\not= 0$ by (53), (54).
Of course $J^{(2)}$ is the energy-momentum tensor which should not have an 
anomalous dimension.

Now we set the dimension $d=3$. From
\begin{eqnarray}
\eta_{1}(\varphi) &=&  \frac{2\sin\pi\mu}{\pi}\frac{\Gamma(2\mu-2)}{\Gamma
(\mu+1)\Gamma(\mu-2)}
\label{55} \\
\eta_{1}(\varphi)|_{d=3} &=& \frac {4}{3\pi^{2}}  
\label{56}
\end{eqnarray}
we obtain
\begin{equation}
\eta_{1}(l) =  \frac{16(l-2)}{3\pi^{2}(2l-1)}
\label{57}
\end{equation}
Unitarity requires this anomalous dimension to be positive which it is 
indeed. Inserted into (7) we obtain the final mass formula
\begin{equation}
m(l)^{2} = \frac {16}{3N\pi^{2}}(l-2) + O(1/N^{2})
\label{58}
\end{equation}

\section{Conclusion}
The result (58) for the mass squared of the gauge bosons at leading order 
of 1/N is surprisingly simple due to cancellations between the five graphs 
primarily. Contrary to the case of the four-point function of the scalar field 
$\alpha$, the four-point function (12) does apparently not reduce to an 
elementary function at $d=3$ (see (34)). The linear dependence of the mass 
squared on the spin $l$ of the boson is of Regge trajectory type and suggests
the existence of a string theory from which the higher spin field theory on 
AdS(4) can be derived.

\end{document}